\documentclass{article}

\usepackage{amsfonts, amsmath, amssymb, amsbsy}
\usepackage{epsfig}
\usepackage{graphicx}
\usepackage{longtable}

\begin{document}

\begin{center}
\textbf{Optical and X-ray behaviour of the high mass X-ray transient A0535+26/HDE245770 in February-March 2014}
\end{center}

\begin{center}
\textbf{F. Giovannelli$^1$, G.S. Bisnovatyi-Kogan$^{2,3}$, I. Bruni$^4$,\\ G. Corfini$^5$, F. Martinelli$^5$, C. Rossi$^6$}
\end{center}

\begin{center}
{\it
\noindent $^1$INAF - Istituto di Astrofisica e Planetologia
Spaziali,\\
Area di Ricerca di Roma-2, Via del Fosso del Cavaliere 100, I
00133 Roma, Italy \\
$^2$Space Research Institute, Profsoyuznaya 84/32,
Moscow 117810, Russia\\
$^3$National Research Nuclear University MEPhI, Kashirskoe Shosse 31,
Moscow 115409, Russia\\
$^4$INAF - Osservatorio Astronomico di Bologna, Stazione astronomica di Loiano,\\Strada della Futa (SS N. 65) -
I 40050 Loiano, Italy\\
$^5$Montecatini Val di Cecina Astronomical Centre - Loc. Palareta, 18\\ I 56040 Montecatini Val di Cecina, Italy \\
$^6$Dept. of Physics, Roma Sapienza University, - P.le Aldo Moro 2 - I 00185 Roma, Italy
}
\end{center}

\begin{abstract}
The optical behavior of the Be star in the high-mass X-ray transient A0535+26/HDE245770 shows that at periastron the luminosity is typically enhanced by 0.02 to a few tenths mag, and the X-ray outburst occurs eight days after the periastron. Indeed, at the periastron an increase of the mass flux occurs. This sort of flush reaches the external part of the temporary accretion disk around the neutron star and moves to the hot central parts of the accretion disk and the neutron star's surface. The time necessary for this way is dependent on the turbulent viscosity in the accretion disk, as discussed  by  Giovannelli, Bisnovatyi-Kogan, and  Klepnev (2013) (GBK13).
In this paper we will show the behaviour of this system in optical band around the predicted periastron passage on 21st February 2014, by using the GBK13's ephemeris
that we used to schedule our spectroscopic and photometric optical observations. Spectroscopic unusual activity detected in the Balmer lines and the enhancement in the emission in B, V, and R bands around the periastron passage, and the subsequent X-ray event definitively demonstrate the existence of a $\sim$ 8-day delay between optical and X-ray flares.

\end{abstract}

\section{Introduction}
The X-ray source A 0535+26 was discovered by the Ariel V satellite on April 14, 1975 (Coe et al., 1975). The X-ray source was in outburst with  intensity of $\sim 2$ Crab) and showed a pulsation with the rate of $\sim 104$ s (Rosenberg et al., 1975). The hard X-ray spectrum during the decay from the April 1975 outburst became softer, so that the 19 May spectrum had $E^{-0.8}$ and the 1 June spectrum $E^{-1.1}$ (Ricketts et al., 1975).

Between 13 and 19 April, 1975, as the nova brightened, the spectra showed some evidence of steepening. The best fit of the experimental data between roughly 27 and 28 April was compatible with an 8 keV black-body curve (Coe et al., 1975). The X-ray source decayed from the outburst with an {\it e}-folding time of 19 days in the energy range of 3-6 keV (Kaluzienski et al., 1975). The Be star HDE 245770 was discovered as the optical counterpart of A 0535+26 by Bartolini et al. (1978), and was classified as an O9.7IIIe star by Giangrande et al. (1980). This is the most reliable classification of this star.

Complete reviews of this system can be found in Giovannelli et al. (1985), Giovannelli \& Sabau-Graziati
(1992), and Burger et al. (1996).

Briefly, the properties of this system, placed at a distance of $1.8 \pm 0.6$ kpc (Giangrande et al., 1980), can be summarized as follows: a hard X-ray transient, long-period X-ray pulsar -- the secondary star -- is orbiting around the primary O9.7IIIe star. The masses are of $\sim 1.5 \pm 0.3$ M$_\odot$ (Joss \& Rappaport 1984; Thorsett et al. 1993; van Kerkwijk, van Paradijs, J. \& Zuiderwijk, 1995), and 15 M$_\odot$ (Giangrande et al., 1980) for the secondary and primary star, respectively. The eccentricity is e = 0.497 (Finger et al., 1994). Usually the primary star does not fill its Roche lobe (de Loore et al., 1984).
However, the suggestion that there might be a temporary accretion disk around the X-ray pulsar when it approaches periastron (Giovannelli \& Zi\'{o}{\l}kowski, 1990) was confirmed by the X-ray measurements of Finger et al. (1996) and was discussed by Giovannelli et al. (2007).

By using two measurements in optical during two identical decays from relative maxima of the luminosity of HDE 245770, Bartolini et al. (1983) determined the orbital period of the system HDE 245770/A 0535+26 as P$_{\rm orb} = 110.856 \pm 0.002$ days. These authors assumed the time of the maximum flare luminosity observed by R\"{o}ssiger (1978) as the reference maximum, that is, JD(L$_{max}) = 2,443,496.37$ + n$\times$110.856.
Thus, Bartolini et al. (1983) obtained for the maximum of the optical flare of December 5, 1981 (JD 2,444,944.5) (Giovannelli et al., 1985) (hereafter 811205-E; E stands for event) the computed time JD 2,444,937.5,  which agrees with the observed time. For this reason Bartolini et al. (1983) gave JD $2,444,944 \pm 10$ as the time of the occurrence of the 811205-E.
But, because the 811205-E is clearly peaked at that date and triggered the subsequent short X-ray outburst of December 13, 1981 (811213-E) (Nagase et al., 1982), Giovannelli \& Sabau-Graziati (2011) assumed the ephemeris of the system as JD$_{\rm opt-outb}$ = JD$_0$(2,444,944) $\pm$ n(110.856 $\pm$ 0.002) days. With such ephemeris they showed a systematic delay between optical and X-ray outbursts after and before 811213-E. However, the time delay between optical and X-ray outbursts, starting from 811205-E, is becoming longer for the most recent outbursts. This suggests that the orbital period determined by Bartolini et al. (1983) that was also  used by Giovannelli \& Sabau-Graziati (2011) is slightly too short.
Therefore Giovannelli, Bisnovatyi-Kogan \& Klepnev (2013) assumed the orbital period determined by Priedhorsky \& Terrell (1983) from X-ray data: P$_{\rm orb} = 111.0 \pm 0.4$ days, and the ephemeris JD$_{\rm opt-outb}$ = JD$_0$(2,444,944) $\pm$ n(111.0 $\pm$ 0.4) days --- the 111-day orbital period agrees within the error bars with many other determinations reported in the literature (from optical data, e.g. Guarnieri et al., 1985; de Martino et al., 1985; Hutchings, 1984; Janot-Pacheco, Motch \& Mouchet, 1987. From X-ray data, e.g. Nagase et al., 1982; Priedhorsky \& Terrell, 1983; Motch et al., 1991; Finger et al., 1996; Coe et al., 2006) --- in order to show the 8-day delay between optical and X-ray outbursts as characteristic of HDE245770/A0535+26, and the construction of a model for explaining this delay.

Assuming that this 8-day delay is effectively a characteristic of this binary system, we scheduled our optical spectroscopic and photometric observations around the passage at the periastron of the neutron star on February 21, 2014 -- the 106th cycle after 811205-E for detecting an optical flare and expecting then the subsequent X-ray outburst.

In this paper we will present our experimental data and a discussion of the results.

\section{Experimental results}

We planned optical photometry and spectroscopy of HDE 245770 at the Loiano observatory with the 1.52 m Cassini telescope, and photometry at the Montecatini Val di Cecina Astronomical Centre 36 cm telescope around the periastron passage of the neutron star A0535+26 (JD 2456710) following the ephemeris of  Giovannelli, Bisnovatyi-Kogan \& Klepnev (2013), that used the orbital period of $111.0 \pm 0.4$ days determined by Priedhorsky \& Terrell (1983). This passage is just 106 orbital periods after the 811205-E. The first results were reported in Giovannelli et al. (2014).

The instrument used in the focus of the Loiano telescope is the BFOSC, whose characteristics can be found at
http://www.bo.astro.it/~loiano/TechPage/bfosceng/
\noindent
BFOSC.html.

For the photometric measurements at Val di Cecina observatory, a CCD SBIG Kaf 1600 has been used in the focus of the Schmidt Cassegrain 36 cm telescope with a V filter.

\subsection{Optical photometry}

Figure 1 shows the results of our B, V and R photometry, together with the X-ray data from the IBIS/ISGRI (18-60 keV) and  JEM-X1 (3-10 keV) (Sguera et al., 2014), and from the MAXI/GSC (4-10 keV) (Takagi et al., 2014). The periastron passage is marked with a red line and corresponds to the 106th cycles after 811205-E.

\begin{figure*}[!ht]
\begin{center}
\includegraphics[width=7.5cm]{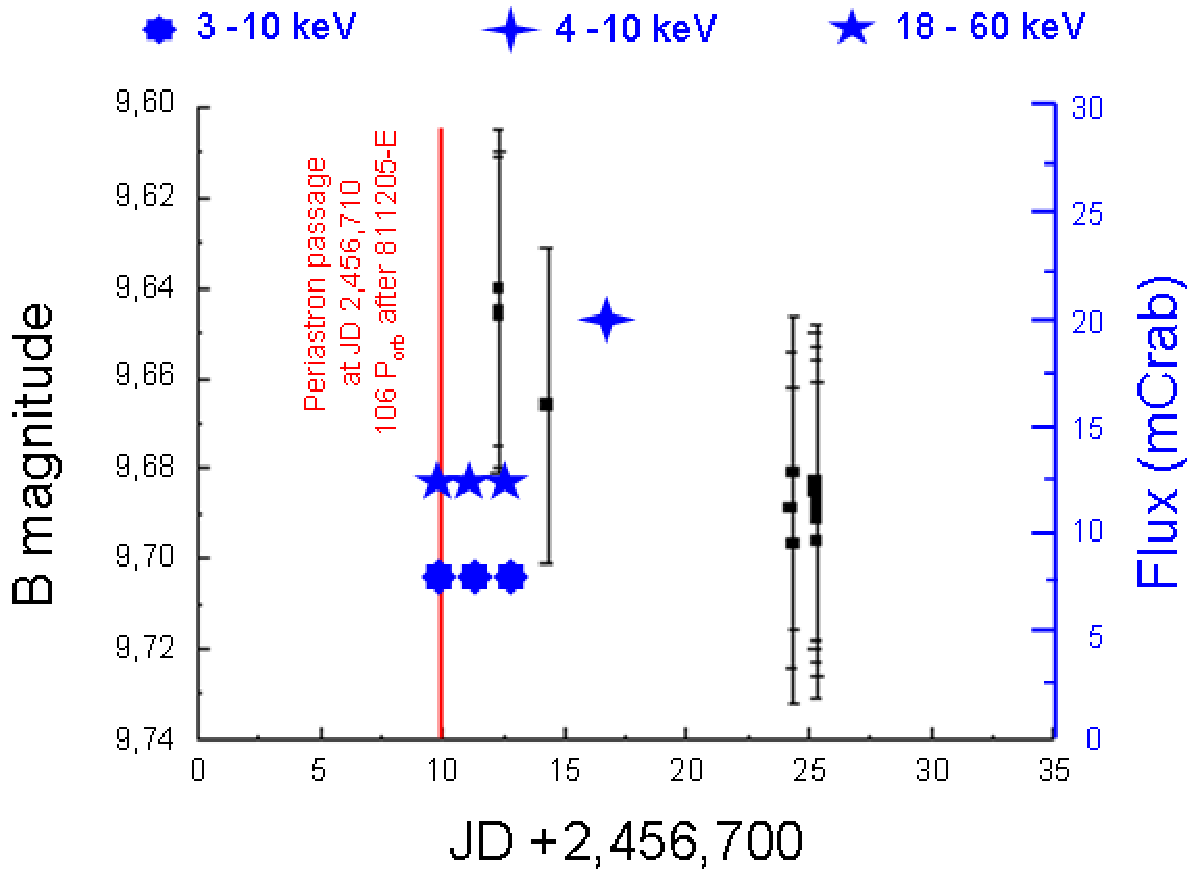}
\includegraphics[width=7.5cm]{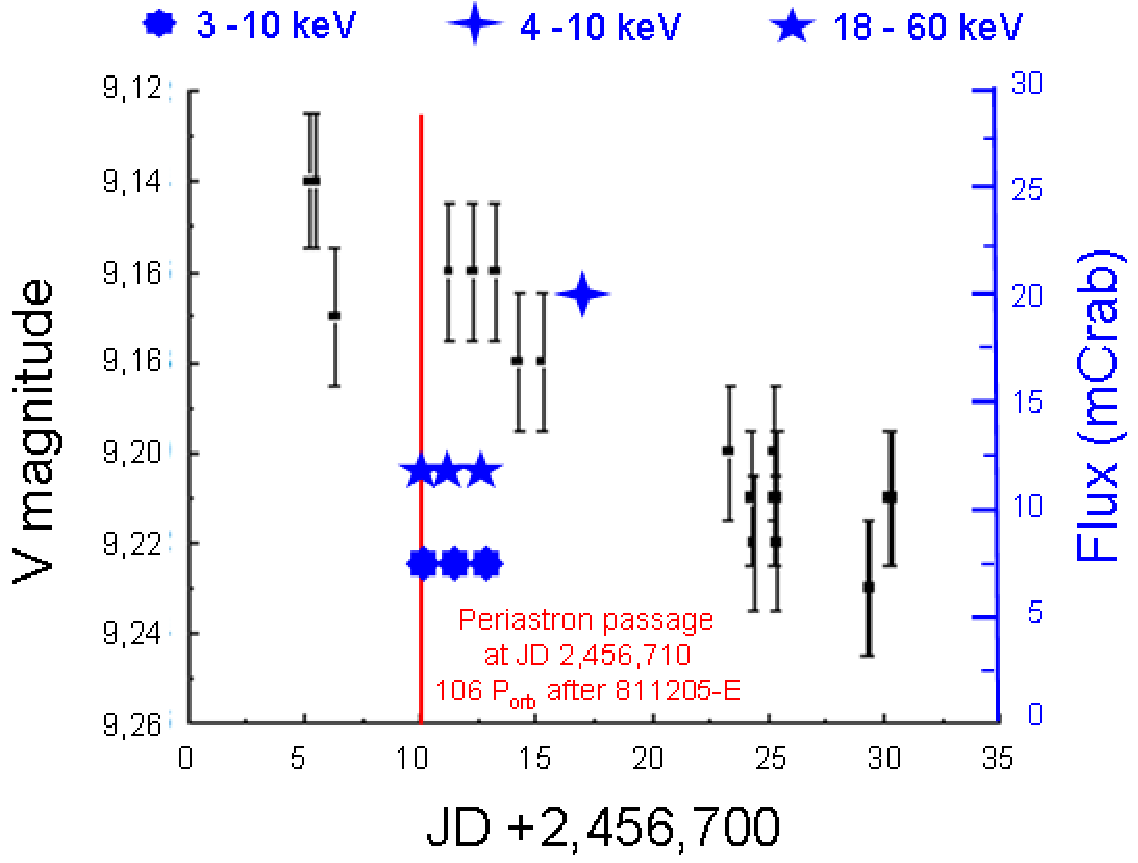}
\includegraphics[width=7.5cm]{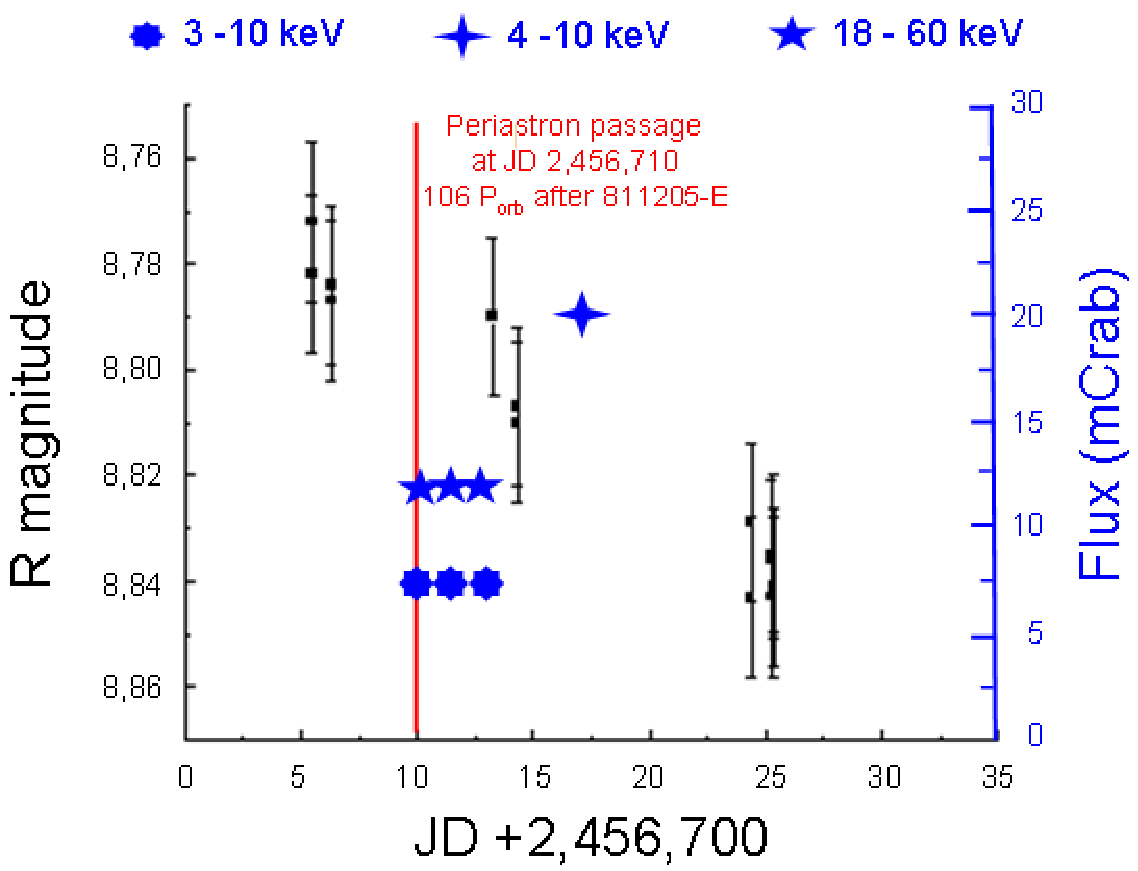}
\caption{Optical outburst on February 2014. X-ray data are marked with blue stars (3-10 keV and 18-60 keV, Sguera et al., 2014; 4-10 keV, Takagi et al., 2014). Optical B, V, and R data from up to down are marked in black. The periastron passage is marked with a red line. It occurred roughly seven days before the maximum of the X-ray outburst.} \label{lc}
\end{center}
\end{figure*}

It is evident that the X-ray outburst follows the periastron passage of about seven days (no more data are available), where the optical B, V, and R emissions are roughly at the relative maximum.

\subsection{Optical spectroscopy}

Spectroscopic measurements allowed us to derive the equivalent widths (EWs) of H$_\alpha$ and H$_\beta$ in emission.
Figure 2 shows the module of the H$_\alpha$ (left) and H$_\beta$ (right) EWs, together with the X-ray data. It is clearly evident a jump in the H$_\alpha$ EW in correspondence with the rise of X-ray intensity.
Also H$_\beta$ EW increases, but its jump seems to be delayed with respect to that of H$_\alpha$.

\begin{figure*}
\begin{center}
\includegraphics[width=7.5cm]{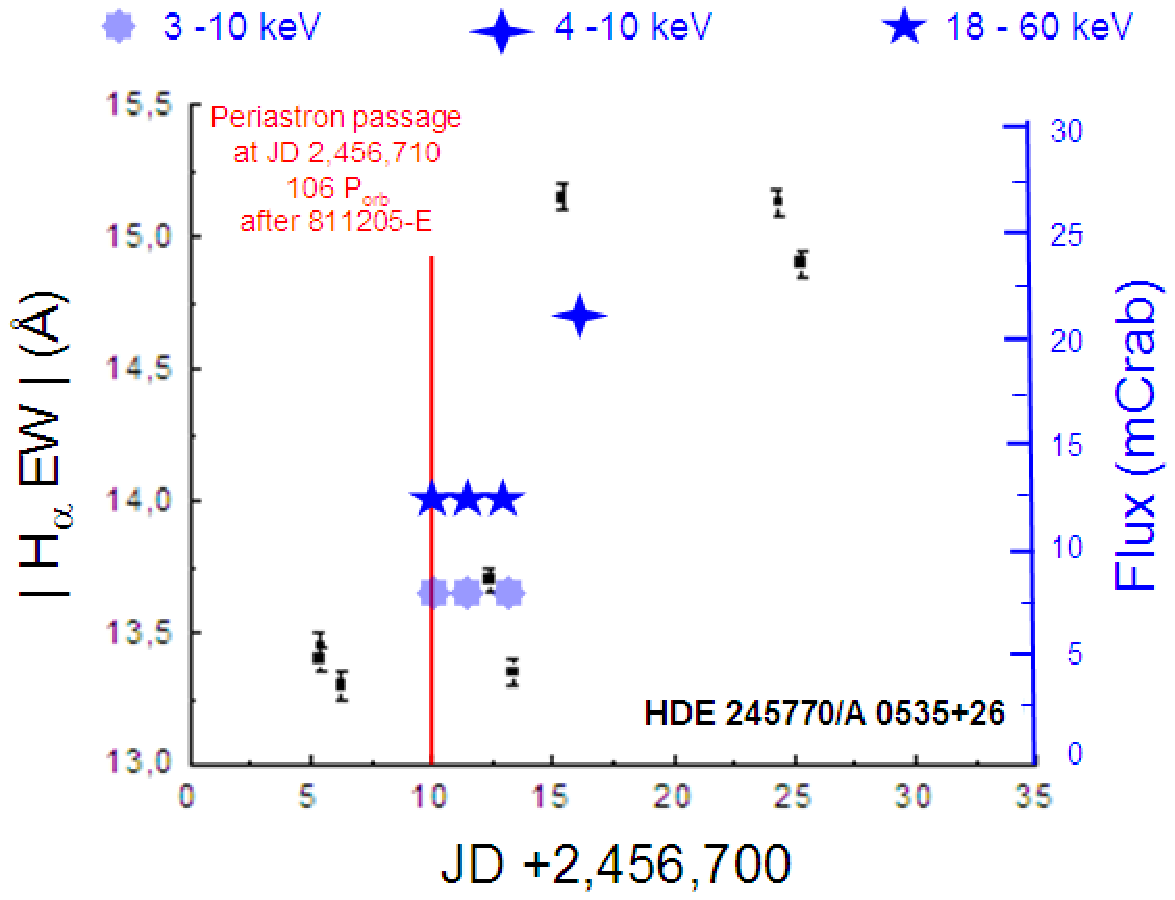}
\includegraphics[width=7.5cm]{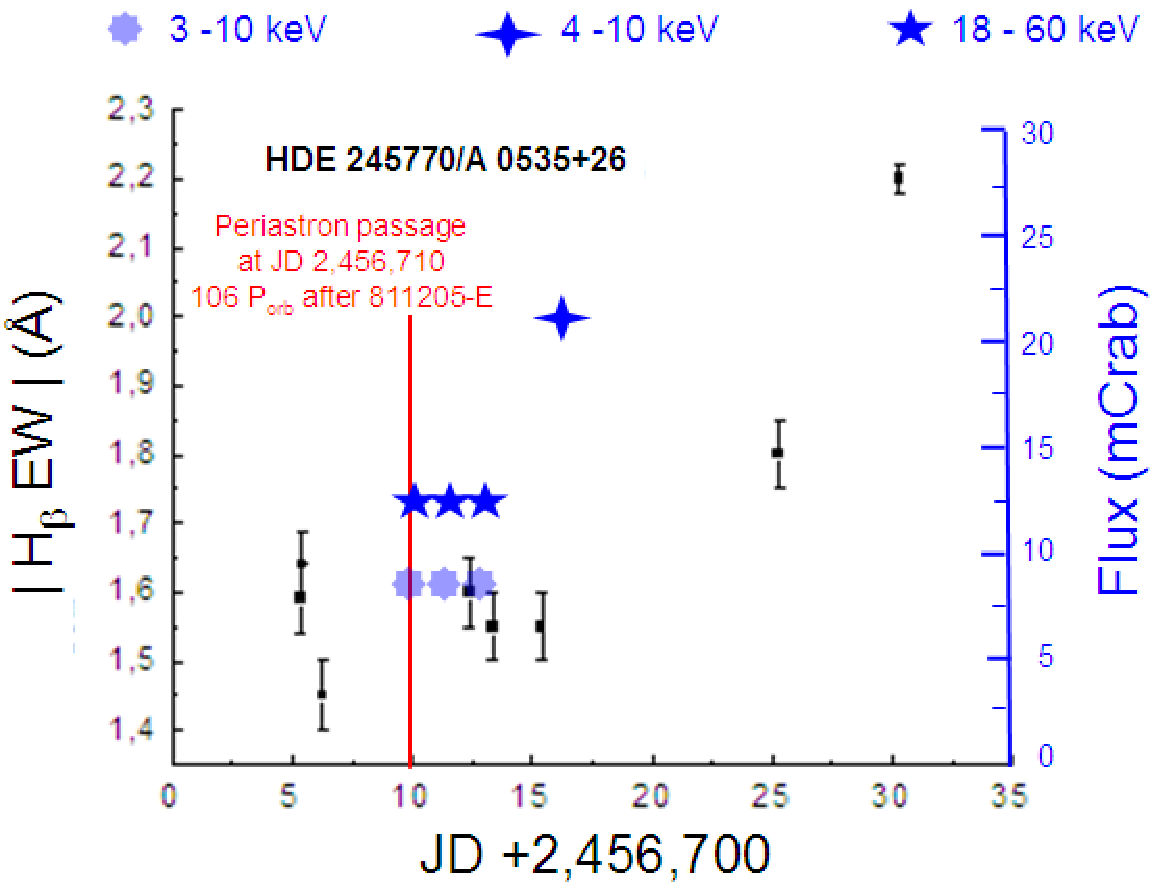}
\caption{Equivalent widths, marked in black, of H$_\alpha$ (up) and H$_\beta$ (down) in emission during the optical and X-ray outbursts on February 2014. X-ray data are marked with blue stars (3-10 keV and 18-60 keV, Sguera et al., 2014; 4-10 keV, Takagi et al., 2014). The periastron passage is marked with a red line. } \label{lc}
\end{center}
\end{figure*}

\subsection{Relationship between optical and X-ray outbursts}

Looking at the literature, we constructed the diagram shown in Fig. 3 where we plotted the intensity of the optical flares, occurred at the periastron passage versus the intensity of the corresponding X-ray outbursts, occurred about eight days after. The best fit shows a linear relationship between the intensity of the X-ray outbursts and the intensity of the optical flares, in agreement with the accretion processes independent of any kind of model, and moreover just in agreement with the quantitative model developed by Giovannelli, Bisnovatyi-Kogan \& Klepnev (2013).

\begin{figure}[!hbp]
\begin{center}
\includegraphics[width=8cm]{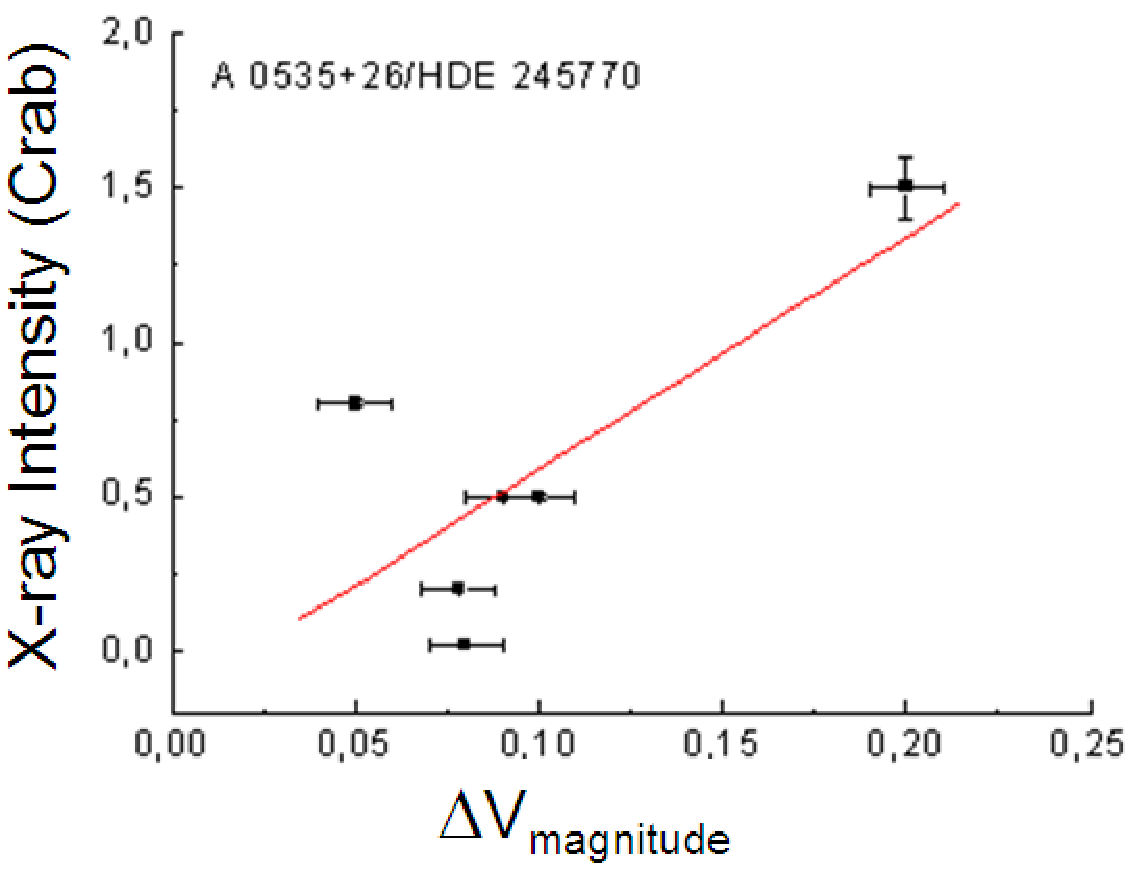}
\caption{Intensity of X-ray outbursts in A 0535+26 versus V magnitude variations in HDE 245770 for different events in which both X-ray and optical data are available  (Giovannelli \& Sabau-Graziati, 1992, Giovannelli, Bisnovatyi-Kogan \& Klepnev, 2013 and references therein, and this work).} \label{lc}
\end{center}
\end{figure}

\section{Discussion and conclusions}


The event of February--March 2014 allowed to confirm definitively that A0535+26/Flavia's star system scans the neutron star passage at periastron with a periodicity P$_{\rm orb} = 111.0 \pm 0.4$ days. The ephemeris we used -- JD$_{\rm opt-outb}$ = JD$_0$(2,444,944) $\pm$ n(111.0 $\pm$ 0.4) days -- was derived from the orbital period of the system (Priedhorsky \& Terrell, 1983) and from the optical flare of December 5, 1981 (811205-E), followed by the X-ray flare of December 13, 1981 (811213-E) (Giovannelli, Bisnovatyi-Kogan \& Klepnev, 2013). Also for this event (106 orbital periods after 811205-E) we have had the confirmation that the passage of A0535+26 at the periastron occurs about eight days before the subsequent X-ray outburst. This fact agrees with the time--delay model developed by Giovannelli, Bisnovatyi-Kogan \& Klepnev (2013).

The jump in the EW of H$_\alpha$ and H$_\beta$ in correspondence with the rise of the X-ray outburst is an important result that deserves further investigation. However, such an increase was detected by Giovannelli, Gualandi \& Sabau-Graziati (2010) soon after the 93rd periastron passage after 811205-E, that anticipated the incoming X-ray outburst (Caballero et al, 2010a,b,c; Caballero et al., 2011) (see Fig. 4). Optical unusual activity was detected also at the 96th periastron passage after 811205-E (Giovannelli, Nesci \& Rossi, 2011).
Also Camero-Arranz et al. (2011) found similar behaviour around several X-ray outbursts.

\begin{figure}
\begin{center}
\includegraphics[width=8cm]{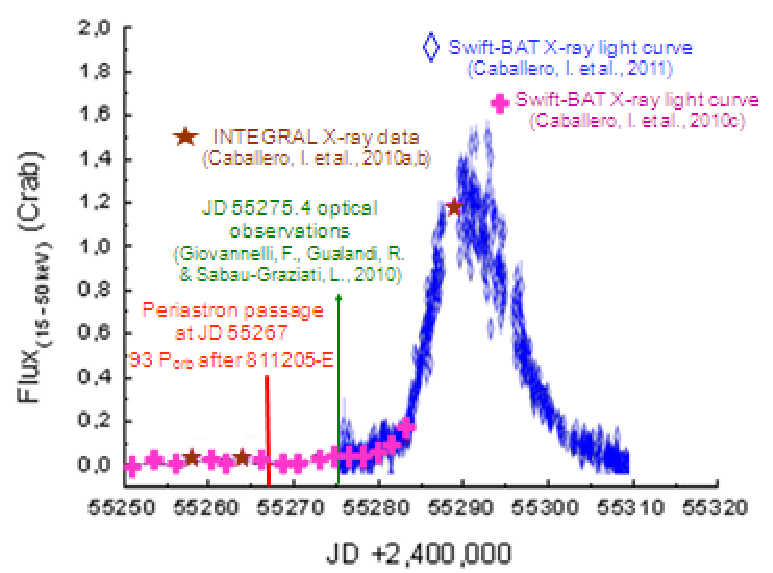}
\caption{X-ray outburst of A 0535+26 after the 93rd periastron passage after 811205-E. Red line shows the periastron passage and the green line marks the epoch of optical spectroscopic measurements. Data and references are clearly shown in the figure.} \label{lc}
\end{center}
\end{figure}

In order to enforce the validity of the ephemeris derived firstly by Bartolini et al. (1983), reconsidered by Giovannelli \& Sabau-Graziati (2011) and refined by Giovannelli, Bisnovatyi-Kogan \& Klepnev (2013) by using the orbital period derived by Priedhorsky \& Terrell (1983), we can consider the 107th periastron passage on June 12, 2014 (JD 2,456,821) after the 811205-E, around which it was not possible to perform optical measurements because of the invisibility of the system having coordinates $\alpha_{1950} = 05^h:35^m:48^s$ and $\delta_{1950} = +26^\circ:17^\prime:18^{\prime\prime}$. An X-ray emission was detected by the FERMI/GMB during the period 10--16 June 2014, corresponding to JD 2,456,819--825 (Finger \& Jenke, 2014).
This fact is a further confirmation of the validity of the ephemeris used.

We have found an experimental linear relationship between the intensity of X-ray outbursts of A0535+26 and the intensity of the optical flares of HDE 245770 that agrees with any kind of model adopted for describing the accretion process around a neutron star in X-ray/Be systems: higher is the mass involved in the accretion process, higher is the intensity of the corresponding X-ray emission.

We strongly encourage optical measurements around the 108th and 109th periastron passage on 2014 October 1 (JD +2,456,932), and on 2015 January 20 (JD +2,457,043). The optimum period for the optical observations should be in the ranges September 10 -- October 20, 2014, and January 1 -- February 10, 2015, possibly with measurements performed each night, at least one daily photometric point, possibly in B, V, R, I bands. This just for detecting the optical flares at the periastron passages. Also spectroscopic measurements for detecting the first three Balmer lines should be desiderable and important for detecting the jump in the equivalent widths in correspondence with the rise of the predicted X-ray outbursts. For this reason, we invite the community of X-ray astronomers to be in alert starting from October 1 till at least 15 October, 2014, and from January 20 till at least February 5, 2015. During those intervals of time, X-ray outbursts should rise rather fast from the "quie
 scent" emission to the peak (reached about eight days after the periastron passage) and a slower decay to "quiescence" as predicted by the model of Giovannelli, Bisnovatyi-Kogan \& Klepnev (2013).

\section{Acknowledgements}
We would like to thank the staff of the Loiano observatory for the allocated time and excellent support for the photometric and spectroscopic observations. A particular thank to the staff of the Montecatini Val di Cecina Astronomical Centre for the allocated time for the photometric observations.

This research made use of NASA's Astrophysics Data System.

 \end{document}